\documentstyle[12pt,epsf]{article}
\begin{document}
\begin{titlepage}
\title{SU(3) Baryon Chiral Perturbation Theory and Long Distance
Regularization$^*$}
\author{John F. Donoghue$^1$, Barry R. Holstein$^{1,2}$,
and Bu\={g}ra Borasoy$^1$\\
$^1$Department of Physics and Astronomy\\
University of Massachusetts\\
Amherst, MA  01003, USA\\
$^2$Institut f\"{u}r Kernphysik\\
Forschungszentrum J\"{u}lich\\
D-52425 J\"{u}lich, Germany}
\maketitle
\begin{abstract}
The use of SU(3) chiral perturbation theory in the analysis of low
energy meson-baryon interactions is discussed.  It is emphasized that short
distance effects, arising from propagation of Goldstone bosons over
distances smaller than a typical hadronic size, are model-dependent
and can lead to a lack of convergence in the SU(3) chiral expansion if
they are included in loop diagrams.
In this paper we demonstrate how to remove such effects in a chirally
consistent fashion by use of a cutoff and demonstrate that such
removal ameliorates problems which have arisen in previous
calculations due to large loop
effects.
\end{abstract}
\vfill
$^*$Research supported in part by the National Science Foundation
and the Deutsche Forschungsgemeinschaft
\end{titlepage}

\section{The Problem}

The low energy phenomenology of baryons is relatively simple. In the 1960's,
this
simplicity was evidenced by the successes of SU(3) symmetry. Indeed, masses and
couplings can be well described by SU(3) invariant interactions with SU(3)
breaking at
the 5 - 25\% level. In the present era we have come to understand this
invariance in terms of QCD and the underlying quark substructure of
baryons---SU(3) relations
work because the effects of the s-u-d mass splittings
are relatively small. Moreover, the quark model even allows us to
understand many details of
the pattern of SU(3) symmetry breaking. Overall, most
features of the static properties of baryons are reasonably well
understood.

It has also been realized that the old SU(3) results represent merely
the lowest order terms of an expansion in energy and quark masses in a
rigorous effective field theory framework which exploits the (broken)
SU(3)$_L\times$SU(3)$_R$ chiral symmetry of the QCD Lagrangian. The
higher order terms in this expansion can be calculated via the
technique called ``chiral perturbation theory'', which has already
been highly developed and successfully applied within the sector of
Goldstone boson interactions.\cite{GL} In the related case of baryon-Goldstone
interactions, there has also been a great deal of activity using
methods generalized from the purely mesonic situation.\cite{HBCPT}

However, the problem is that traditional SU(3) baryon chiral perturbation
theory
does not appear to work well.  As generally applied, it does not manifest the
approximate SU(3) symmetry that one sees in the real world, in that SU(3)
breaking corrections in loop diagrams often appear at the 100\%
level. It is particularly distressing that these effects come from the most
apparently model-independent parts of the theory---the nonanalytic chiral
loops.
With some parameter fitting, it appears in practice that 
such effects can be compensated by
positing equally large effects from the effective Lagrangian at higher
order in the chiral expansion. However, this leads to worries about
convergence. In any event, the simplicity evident in baryon
physics has become lost. In its conventional manifestation then, SU(3)
baryon chiral perturbation theory does not represent a good first
approximation to baryon physics.

In this paper we will suggest a resolution to this problem in terms of a
reformulation of baryon chiral perturbation theory within a framework which is
better suited to phenomenological applications. Before we turn to a diagnosis,
let us, however, demonstrate the nature of the
problem by observing several pertinent results. In each case we defer
the specifics of the chiral analysis until later in the paper and
simply quote results in order to convince the reader that a problem
exists.

\begin{itemize}

\item [i)] Baryon masses can be understood by noting that the
quark mass nondegeneracy arises from a component of ${\cal L}_{\rm
QCD}$ which can be represented in terms of a Lorentz scalar SU(3) octet
$\bar{q} \lambda_8 q$ operator.  To first order in symmetry breaking one can
then
write the
baryon octet masses in terms of an SU(3) invariant term $\hat{M}_0$ plus
octet $f_m,d_m$ couplings---
\begin{eqnarray}
M_N&=&\hat{M}_0-4m_K^2d_m+4(m_K^2-m_\pi^2)f_m\nonumber\\
M_\Lambda&=&\hat{M}_0-{4\over 3}(4m_K^2-m_\pi^2)d_m\nonumber\\
M_\Sigma&=&\hat{M}_0-4m_\pi^2d_m\nonumber\\
M_\Xi&=&\hat{M}_0-4m_K^2d_m-4(m_K^2-m_\pi^2)f_m\label{eq:cc}
\end{eqnarray}
Since the four octet baryon masses are represented in terms of
effectively three parameters there is a corresponding sum rule---that
of Gell-Mann and Okubo\cite{GMO}---
\begin{eqnarray}
M_\Sigma-M_N &=& {1\over 2}(M_\Xi-M_N)+{3\over
4}(M_\Sigma-M_\Lambda)\nonumber\\
{\rm Expt.}:\quad 254 {\rm MeV}&=&248 {\rm MeV}
\end{eqnarray}
which is satisfied experimentally at the $3\%$ level.

When analyzed in the usual fashion in 
chiral perturbation theory, however, this simplicity
is lost. At one loop---${\cal O}(q^3)$---order the chiral loop corrections are
found to
be extremely large\cite{CHIM}
\begin{eqnarray}
&& \delta M_N=-0.31\:{\rm GeV};\quad\delta M_\Sigma=-0.67\:{\rm GeV};
\nonumber \\
&&\delta M_\Lambda=-0.66\:{\rm GeV};\quad\delta M_\Xi=-1.02\:{\rm GeV}
\end{eqnarray}
such that, {\it e.g.}, the $\Xi$ mass receives a 100\%
correction.  This calculation has also been carried out to---${\cal O}(q^4)$---by
Borasoy and
Meissner\cite{BM}, who quote their results as
\begin{eqnarray}
M_N&=&\bar{M}(1+0.34-0.35+0.24)\nonumber\\
M_\Sigma&=&\bar{M}(1+0.81-0.70+0.44)\nonumber\\
M_\Lambda&=&\bar{M}(1+0.69-0.77+0.54)\nonumber\\
M_\Xi&=&\bar{M}(1+1.10-1.16+0.78)\label{eq:aa}
\end{eqnarray}
where the non-leading terms above refer to the contribution from
${\cal O}(q^2)$ counterterms, nonanalytic pieces of ${\cal O}(q^3)$,
and ${\cal O}(q^4)$ counterterms respectively.  Obviously,
the contribution from higher order terms is far larger than one expects
and the series does not display obvious convergence. Also, the Gell-Mann-Okubo
deviation is found to be five times larger than experiment.

\item [ii)] Baryon axial couplings can be related by noting
that the weak axial current arises from an SU(3) octet
$\bar{q}'\gamma_\mu\gamma_5 q$ structure.  Thus to leading order in
SU(3) the various weak matrix elements can be represented in terms of
simple $f_A,d_A$ couplings. A fit  to the ten experimentally
measured semileptonic hyperon decay
rates is found to yield reasonable results, with $\chi^2/$ d.o.f. $\sim$ 1.
SU(3) breaking in the decay rates is noticeable, but the amount of
SU(3) breaking is never above 5\%.\cite{DHS}
One can explore quark models and
find that they generate breaking that is of about this magnitude, and
the challenge then is to fit the pattern of breaking.

When chiral loops are calculated,\cite{BSW} one finds logarithmic dependence
on the meson masses that leads to significant SU(3) breaking. Typically these
effects are too large.
Numerically, choosing a renormalization scale $\mu\sim 1$ GeV, typical
leading log corrections are found to be at the 30-50\% level and a fit
to the experimental hyperon decay rates finds a much increased
chi-squared---the chiral corrections go in the wrong direction!

\item [iii)] S-wave nonleptonic hyperon decay amplitudes can be
related by using the feature that the octet component of the weak
Hamiltonian is dominant over its 27-dimensional counterpart by a
factor of twenty or so, plus using chiral symmetry to relate the
experimental pion decay amplitudes to simpler baryon to baryon matrix
elements.  This allows a fit in terms of octet $f_w,d_w$ parameters:
Such a representation yields a very good fit to the experimental
amplitudes in that the two independent predictions:\footnote{Note that
the second of these results is the Lee-Sugawara sum rule.\cite{LS}}
\begin{eqnarray}
A(\Sigma^+_+) &=& 0\quad {\rm vs.}\,\, 0.13\times 10^{-7}\,\,({\rm
expt.})
\nonumber\\
\sqrt{3}A(\Sigma^+_0) &-& 2A(\Xi^-_-)-A(\Lambda^0_-)=0\quad {\rm
vs.}\,\,0.11\times 10^{-7}\,\,({\rm expt.})
\end{eqnarray}
are, since the typical size of an s-wave amplitude is $\sim 4\times
10^{-7}$, both reasonably well satisfied by the data.
\footnote{It is, of course,
possible to apply a similar analysis to the corresponding P-wave
amplitudes.  However, in this case the leading piece of each amplitude
involves a significant cancellation from from pairs of baryon pole
diagrams, so that there is large and very model dependent sensitivity
to higher order chiral contributions.  Thus we do not analyze this case.}

In baryon chiral perturbation theory, the chiral loop
corrections to individual terms are found to be at the 30-50$\%$
level,\cite{BSW} and a large correction to the Lee-Sugawara relation is found
\begin{equation}
\sqrt{3}A(\Sigma^+_0)-2A(\Xi^-_-)-A(\Lambda^0_-)
\approx -6.4\times 10^{-7}
\end{equation}
which is in considerable disagreement with the experimental number.
The other lowest order prediction---$A(\Sigma^+_+)=0$---is
not affected by chiral logarithms.

One can also see the problem with chiral convergence of individual
terms.  Indeed, a comprehensive analysis of the problem up to second
order counterterms has given\cite{BH}
\begin{eqnarray}
A(\Lambda^0_0)&=&2.35(1+0.62-0.65)\times 10^{-7}\nonumber\\
A(\Sigma^+_0)&=&3.09(1+0.30-0.32)\times 10^{-7}\nonumber\\
A(\Sigma^+_+)&=&0\times 10^{-7}\nonumber\\
A(\Xi^0_0)&=&3.06(1+0.40-0.36)\times 10^{-7}
\end{eqnarray}
where the various contributions are from lowest order, nonanalytic
components, and next order counterterms respectively.

\item [iv)] Hyperon magnetic moments can be related to one
another since they arise from an SU(3) octet $\bar{q}'\gamma_\mu q$
structure. Then to leading order the moments can be written in terms
of simple $f_\mu,d_\mu$ couplings---
\begin{eqnarray}\label{eq:yy}
\mu_p&=&\mu_{\Sigma^+}={1\over 3}d_\mu+f_\mu\nonumber\\
\mu_n&=&2\mu_\Lambda=\mu_{\Xi^0}=-{2\over 3}d_\mu\nonumber\\
\mu_{\Sigma^-}&=&\mu_{\Xi^-}={1\over 3}d_\mu-f_\mu\nonumber\\
\mu_{\Lambda\Sigma}&=&\sqrt{3}\mu_{\Sigma^0}={1\over \sqrt{3}}d_\mu
\end{eqnarray}
The experimental moments are in approximate (but not outstanding)
agreement with these predictions. (Although it is not relevant for our
considerations here we note that the heavier mass, and hence smaller
magnetic moment, of the strange quark explains most of the observed
SU(3) breaking.)

Again the chiral corrections are large and harmful.
Numerically, picking a renormalization scale $\mu=1$ GeV, the
nonanalytic corrections are at the 50-90$\%$ level, and make enormous
modifications of the lowest order results.  As shown by Caldi and
Pagels, there remain three relations, which are independent
of these corrections and are in fact reasonably well satisfied by the
experimental numbers:\cite{CP}
\begin{equation}
\mu_{\Sigma^+}=-2\mu_\Lambda-\mu_{\Sigma^-},\,\,\mu_{\Xi^0}+\mu_{\Xi^-}
+\mu_n=2\mu_\Lambda-\mu_p,\,\,\mu_\Lambda-\sqrt{3}\mu_{\Lambda\Sigma}
=\mu_{\Xi^0}+\mu_n
\end{equation}
However, other relations pose significant problems for experimental
agreement.  Meissner and Steininger have performed a ${\cal O}(q^4)$
analysis of the problem and have shown that it is possible to get good
agreement via a careful choice of counterterms.\cite{MS} The convergence of
the chiral expansion is again a possible problem, as the contributions of terms
of successive orders is found to be
\begin{eqnarray}
\mu_p&=&4.69(1-0.57+0.16)=2.79\nonumber\\
\mu_n&=&-2.85(1-0.36+0.03)=-1.91\nonumber\\
\mu_{\Sigma^+}&=&4.69(1-0.72+0.24)=2.46\nonumber\\
\mu_{\Sigma^0}&=&1.43(1-0.93+0.38)=0.65\nonumber\\
\mu_{\Sigma^-}&=&-1.83(1-0.41+0.04)=-1.16\nonumber\\
\mu_{\Lambda\Sigma}&=&2.47(1-0.57+0.18)=1.51\nonumber\\
\mu_{\Xi^0}&=&-2.85(1-0.95+0.39)=-1.25\nonumber\\
\mu_{\Xi^-}&=&-1.83(1-0.57+0.18)=-0.65
\end{eqnarray}

\end{itemize}

We see in each case then that the chiral corrections are large and in
each situation the leading nonanalytic components destroy the good
experimental agreement which exists at lowest order. There is
something clearly ineffective about this procedure. For a technique
that has aspirations of rigor, this is a dismaying situation
We will show below that the problem resides in a spurious short-distance
contribution that appears in loop diagrams when they are regularized
dimensionally. We propose that we should keep only the long distance
parts of the loops, and propose a cutoff regularization that
accomplishes this.

\section{Effective Field Theory: Separating Long and Short Distances}

Effective field theory is a technique for describing the low energy
limit of a theory. It is an ``effective'' description because it uses
the degrees of freedom and the interactions which are correct at low
energy. All the features of the high energy portion of the theory
are captured in the parameters of a general local effective Lagrangian
which describes the low energy vertices. Using these interactions one
treats the low energy dynamics in a complete field theoretic
description.

Within such a treatment, one encounters loop diagrams, in which the
integration over the momenta includes both low energy and high energy
components. While the low energy portion is fully correct within the
effective theory, the high energy portion is not. One might worry then
about the inclusion of such incorrect high-energy/short-distance
physics present in loops. However, this is not a problem in general
since this high energy effect has the same structure as the terms in
the general local Lagrangian, meaning that any incorrect loop contribution
can be compensated by a shift of the parameters of the Lagrangian. As an
example, the ultraviolet divergences in the effective theory are all
absorbed by defining renormalized parameters.

In practice, there is a situation where such loop effects {\it can} cause
problems. This occurs if the residual short distance contributions are
large even {\it after} renormalization. A large and incorrect short
distance effect 
can still be removed by the adjustment of parameters, but
those parameters must consequently also be large. We then obtain an
expansion which is of the form
\begin{equation}
{\it M} \sim {\it M}_0 ( 1 - 1 + 1 - 1 + ....)
\end{equation}
where each term in the expansion is sizeable and there is no clear
convergence. If one were able to carry out the process to all orders,
one would, of course, still get the right answer. 
However, at any finite order,
the incorrect short-distance phyiscs in loops has obscured the answer
and the expansion is useless. While not formally ``wrong'', this
procedure is ineffective, which is certainly a poor trait for an
effective field theory.

In SU(3) baryon chiral perturbation theory, exactly this situation
occurs when the theory is regularized dimensionally. We will show that
the poor convergence described in the introduction follows largely from
the short-distance component of loop diagrams. In order to provide a
more effective description, we will then reformulate the theory using a
cutoff which retains only the reliable---long-distance---portion of loop
diagrams. This will result in improved phenomenology. Baryon 
effective
field theory becomes even more effective with a long-distance
regularization scheme!

In baryon chiral perturbation theory, the transition between short and
long distance occurs around a distance scale of $\sim$1 fermi, or a momentum
scale of $\sim$200 MeV. This corresponds to the measured size of a baryon
and we will refer to it as the separation scale. The effective field
theory treats the baryons and pions as point particles. This is
appropriate for the very long distance physics - the ``pion tail'' is
independent of whether the baryon is treated as a point particle or an
extended object. However, for propagation 
at distances less then the separation
scale, the point particle theory does not provide
an accurate representation of
the physics - the composite substructure becomes manifest below this
point.

In the next section we focus on the specific Feynman integrals that
arise in baryonic calculations. Our goal is to understand the
structure of
loops in this effective field theory by separating the
short-distance and long-distance physics within the loop integral. The
use of a cutoff representing the separation scale 
will allow us to show that the long distance physics
is well behaved, and that dimensional regularization in practice
contains large short distance contributions for these particular
integrals.

\section{Anatomy of Feynman integrals}

We begin by performing an autopsy on a particular Feynman integral that
appears in the baryon mass analysis. Consider
the
integral
\begin{equation}
\int{d^4k\over (2\pi)^4}{k_ik_j\over
(k_0-i\epsilon)(k^2-m^2+i\epsilon)}
=-i\delta_{ij}{I(m)\over 24\pi} \label{eq:pp}
\end{equation}
where the right hand side simply defines the function $I(m)$.
When regularized dimensionally this has the value
\begin{equation}
I_{dim-reg}(m) = m^3
\end{equation}
This integral is uniquely the source of nonanalytic
corrections to baryon masses.

Some comments about the dimensionally regularized form are instructive.
\begin{itemize}

\item[i)] The Feynman integral is cubicly divergent at high energy.
However a peculiarity of the dimensionally regularized form is that the
result is finite. This is not a problem and occurs at other times in
dimensional regularization. However, it is one indication that
this regularization scheme implies a particular short distance subtraction,
which will in general leave behind finite effects from short distance.
\item[ii)] The only scale in the integral is the meson mass $m$.
Therefore the relevant momenta in the integral all scale with $m$
also. In the limit that $m$ is very large, all of the relevant
momenta correspond to high-energy/short-distance. This is an
indication that as $m$ grows the dimensionally regularized integral
becomes totally dominated by short distance physics---below the separation
scale.
\item[iii)] If we are interested in {\it only} the long distance component of
the integral, this portion would fall off with increasing mass. At
large $m$ the meson progagator could be approximated by a constant
(e.g., as we do for the W- boson mass in low energy weak interactions)
and the low energy portion of the integral would fall as $1/m^2$.
\item[iv)] We would expect that the long distance portion of the
integral would be largest for the smallest meson masses, and greatest
for masssless Goldstone bosons. However the form Eq. \ref{eq:pp} vanishes for
massless particles and is very small for small meson masses.
\end{itemize}

These are all indications that an overall subtraction has taken place which
confuses short and long distance physics. We cannot count on the
dimensionally regularized form to yield only long distance physics---an
implicit short-distance contribution is carried along also.

Now let us isolate the long distance component of the integral.
Indeed it is possible to remove the 
short distance portion by use of a cutoff regularization, as we
demonstrated in ref. \cite{DH}.  Although an exponential cutoff in
three-momentum was employed therein, for our purposes it is most convenient to
employ a simple dipole regulator
\begin{equation}
({\Lambda^2\over \Lambda^2-k^2})^2\label{eq:ff}
\end{equation}
since it enables loop integration to be carried out in terms of simply
analytic forms.  However, the specific shape of the cutoff is
irrelevant---a consistent chiral expansion can always be carried out
to the order we are working.

The introduction of the dipole cutoff Eq. \ref{eq:ff} yields
\begin{equation}
\int{d^4k\over (2\pi)^4}{k_ik_j\over
(k_0-i\epsilon)(k^2-m^2+i\epsilon)}({\Lambda^2\over
\Lambda^2-k^2})^2=-i\delta_{ij}{I_\Lambda
(m)\over 24\pi}
\end{equation}
where
\begin{equation}
I_\Lambda (m)= {1\over2} \Lambda^4 \frac{2m +\Lambda}{(m +\Lambda)^2}
\label{eq:bb}
\end{equation}
Various comments on this form are appropriate
\begin{itemize}
\item[i)] This integral is plotted in Figure 1 for $\Lambda = 400$MeV,
along with its
dimensionally regularized analog. We see that the cutoff result is much
smaller than that of dimensional regularization for kaon and eta
masses. Moreover, what matters for SU(3) breaking are {\it differences}
in the intgral between pions kaons and etas, since a constant effect
can be absorbed into chiral parameters. This difference is quite small
for the cutoff version. We conclude that most of the dimensonally
regularized Feynman integral
for kaons
and etas corresponds to short distance physics.

\item[ii)] The greatest contribution at long distance is seen in the cut-off
scheme
to come from massless mesons, as expected. As the meson mass increases, there
is
a decreasing effect from the long distance portion of the integral.
\item[iii)]We observe then that in the small mass limit
\begin{equation}
I(m)\stackrel{m<<\Lambda}{\longrightarrow}{1\over
2}\Lambda^3-{1\over 2}\Lambda m^2 +m^3+\ldots\label{eq:ee}
\end{equation}
{\it i.e.}, $I(m)$ reduces to the dimensional regularization
result---$m^3$---plus
polynomial terms in $\Lambda$ which are absent in the dimensional
approach. In the next section, we will see explicitly how these polynomial
terms
can be absorbed in the renormalization of chiral parameters.
\item[iv)] In the opposite limit of a large mass compared to the
cutoff
\begin{equation}
I(m)\stackrel{\Lambda<<m}{\longrightarrow}{\Lambda^4\over
m}-{3\over 2}{\Lambda^5\over m^2}+\ldots
\end{equation}
the function $I(m)$ is found to depend upon the pseudoscalar mass to
inverse powers, meaning that the pion will contribute much more than
its heavier eta or kaon counterparts, as we expect intuitively.
\end{itemize}
\vskip 0.5cm

\begin{center} 

\begin{figure}[bth]
\unitlength0.05in  
\begin{picture}(100,75)  
\put(0,0){\makebox(100,75)
{\epsfbox{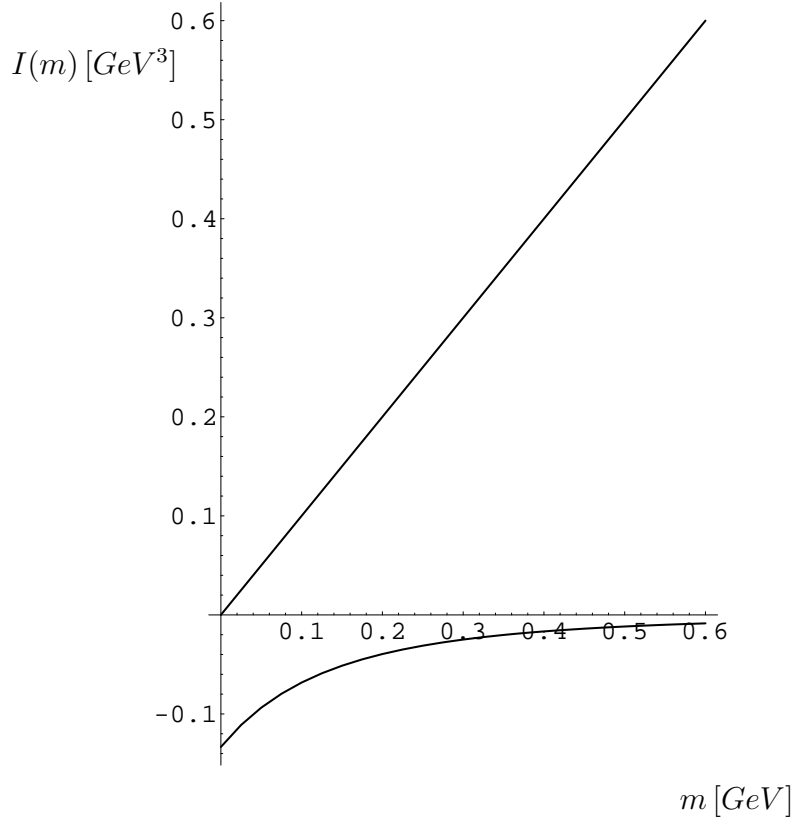}}} 
\put(5,70){$I(m) \, [GeV^3]$}  
\put(75,-7){$m \, [GeV]$} 
\end{picture}
\vskip 1.0cm
\caption{The integral $I(m)$ for the case of dimensional regularization
         ($I=m^3$) and in the cutoff scheme with $\Lambda=400$ MeV.} 
\end{figure}
  
\end{center}

Our conclusion from studying the integral Eq. \ref{eq:pp} is that the cut-off
scheme picks out the long distance part of the integral, which behaves
as expected on physical grounds. In contrast, the dimensional form
carries with it implicit and large contributions from short distance
physics. It is not suprising then that the large short distance
effects 
dominate the analysis when dimensional regularization
is employed and we will demonstrate this explicitly in the next sections.

Before returning to the physics, we analyze the other Feynman integrals
which arise in the analysis of baryon physics.  In the case of baryon axial
couplings and s-wave hyperon decay the relevant heavy baryon integral which
generates the nonanalytic terms in $m^2\ln m^2$ is
\begin{equation}
\int{d^4k\over (2\pi)^4}{k_ik_j\over
(k_0-i\epsilon)^2(k^2-m^2+i\epsilon)}=-i\delta_{ij}{J(m^2)\over 16\pi^2}
\end{equation}
In dimensional regularization the integral has the value
\begin{equation}
J_{dim-reg}(m^2) = m^2 \ln {m^2\over \mu^2}
\end{equation}
while the cutoff version is given by
\begin{equation}
\int{d^4k\over (2\pi)^4}{k_ik_j\over
(k_0-i\epsilon)^2(k^2-m^2+i\epsilon)}({\Lambda^2\over
\Lambda^2-k^2})^2=-i\delta_{ij}{J_\Lambda (m^2)\over 16\pi^2}
\end{equation}
with
\begin{equation}
J_\Lambda (m^2)={\Lambda^4m^2\over (\Lambda^2-m^2)^2}\ln {m^2\over
\Lambda^2}+{\Lambda^4\over \Lambda^2-m^2}
\end{equation}
We plot these forms in Fig. 2. The behavior is qualitatively
similar to that which occured with the previous integral---the dimensional form
overstates the amount of SU(3) breaking. In
addition the growth in the magnitude of the integral at large
masses indicates that short distance physics dominates the dimensionally
regulated form.

\begin{center} 

\begin{figure}[bth]
\unitlength0.05in  
\begin{picture}(100,75)  
\put(0,0){\makebox(100,75)
{\epsfbox{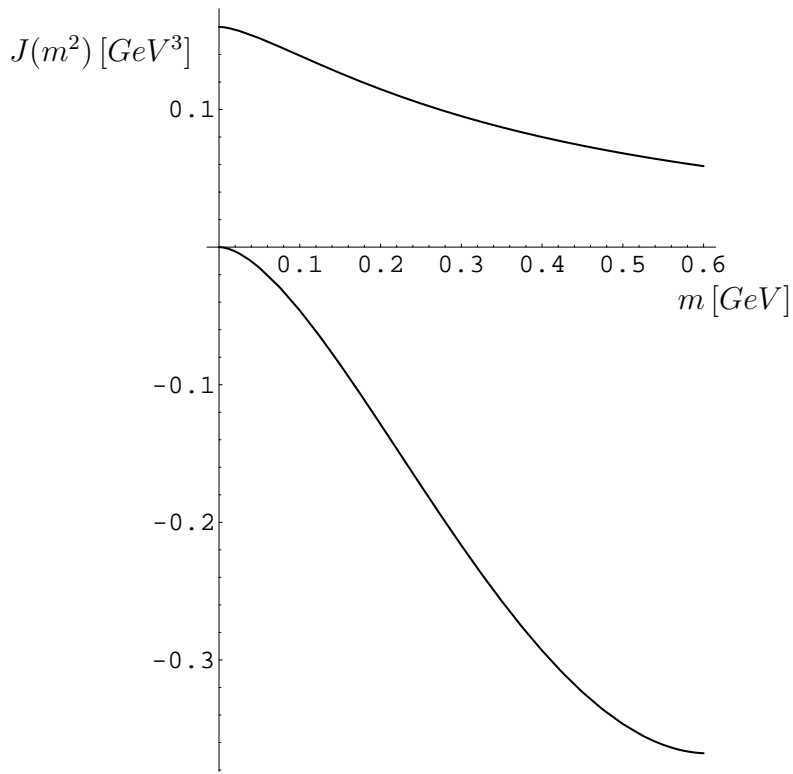}}} 
\put(5,72){$J(m^2) \, [GeV^3]$}  
\put(75,46){$m \, [GeV]$} 
\end{picture}
\vskip 1.0cm
\caption{The integral $J(m^2)$. The lower curve is the result in dimensional 
         regularization, whereas the upper curve shows the case of
         the cutoff scheme with $\Lambda=400$ MeV.} 
\end{figure}
  
\end{center}

The small and large mass limits of the cut-off form are given by
\begin{equation}
J(m^2)\stackrel{m^2 <<\Lambda^2}{\longrightarrow}\Lambda^2+m^2\ln
{m^2\over \Lambda^2}+\ldots
\end{equation}
and
\begin{equation}
J(m^2)\stackrel{m^2>>\Lambda^2}{\longrightarrow}{\Lambda^4\over
m^2}\ln {m^2\over \Lambda^2}+\ldots
\end{equation}
so that again our intutitive expectations are met.

Finally, we consider the integral which is relevant in the analysis
of the magnetic moments
\begin{equation}
\int{d^4k\over (2\pi)^4}{k_ik_j\over
(k_0-i\epsilon)(k^2-m^2+i\epsilon)^2}=
-i\delta_{ij}{K(m)\over 16\pi}
\end{equation}
The dimensionally regularized form is given by
\begin{equation}
K_{dim-reg}(m) = m  \ \ .
\end{equation}
Once again, the integral shows no sign of its true linear divergence, and
grows at large values of $m$, indicating short distance dominance at
large $m$.  The use of the dipole cutoff yields
\begin{equation}
K(m)= -{1\over3} \Lambda^4 \frac{1}{(\Lambda + m )^3} \quad ,
\end{equation}
which is plotted in Fig. 3 and is there compared to
the dimensionally regularized form. Again we see that
the long distance portion of the integral is well behaved and that
dimensional regularization overstates the SU(3) breaking in the
integral.
The function $K(m)$ has the small and large mass limits
\begin{equation}
K(m)\stackrel{m<<\Lambda}{\longrightarrow}-{1\over 3}\Lambda +m+\ldots
\end{equation}
and
\begin{equation}
K(m)\stackrel{m>>\Lambda}{\longrightarrow} -{\Lambda^4\over 3 m^3}+\ldots
\end{equation}
which have the expected qualitative forms.

\begin{center} 

\begin{figure}[bth]
\unitlength0.05in  
\begin{picture}(100,75)  
\put(0,0){\makebox(100,75)
{\epsfbox{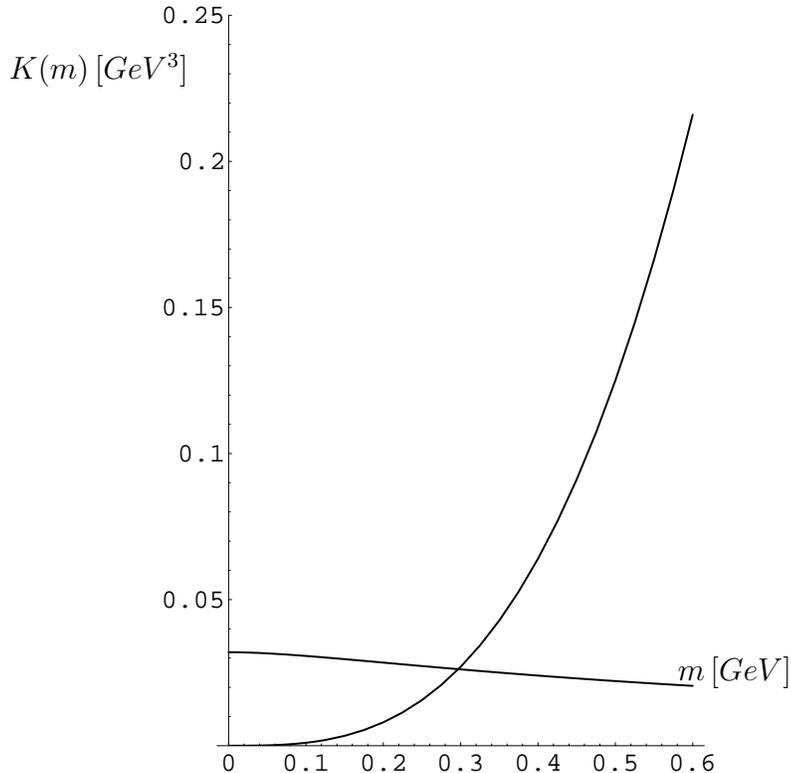}}} 
\put(5,70){$K(m) \, [GeV^3]$}  
\put(75,7){$m \, [GeV]$} 
\end{picture}
\vskip 1.0cm
\caption{The integral $K(m)$. The upper curve is the result in dimensional 
         regularization, whereas the lower one shows the case of
         the cutoff scheme with $\Lambda=400$ MeV.} 
\end{figure}
  
\end{center} 

\section{Theory and Phenomenology with a Cutoff}

In the previous section, we used a cutoff as a tool to explore the
long-distance portion of loop integrals. Since we find the
long-distance portion to be well behaved, we suspect that the problems
described in the introduction are in fact caused by the spurious
inclusion of short-distance effects. We then turn to a different use
for the cutoff - as a regularization technique for handling loop
integrals. 

Field theories can be applied with a variety of regularization
methods. In the end, the resulting 
physics should be independent of the
choice of regularization scheme. At first sight this suggests that it
is unlikely that simply employing 
a change in regularization can have any impact on the
problems mentioned in the introduction. However, we will see that the
choice of a cutoff with a value around the separation scale will
amount to a partial resummation of the chiral expansion and that 
this can be
done without losing the generality of the effective field theory
treatment. If we are right in our assesment that the problem is caused
by spurious loop effects below the separation scale, this resummation
can then lead to an improved procedure for phenomological
applications.

We will first 
show explicitly how the standard chiral expansion is exactly
reproduced for small values of the meson masses. A key ingredient of
this demonstration is the renormalization of the chiral parameters.
The loop integrals will often depend strongly on the value of the
cutoff, and we will encounter integrals with $\Lambda^3$, $\Lambda^2$,
$\Lambda$ and  $ln \Lambda$ dependences, where $\Lambda$ represents the
momentum space cutoff.  However, this does not mean that the resulting
physics will depend on the cutoff this strongly.  Indeed, the final
physics is independent of $\Lambda$.  
This occurs because the terms in
$\Lambda$ have a chiral SU(3) dependence which is the same as the
various terms in the effective Lagrangian. Therefore, in physical
processes one can absorb this $\Lambda$ dependence into a renormalized
value of these parameters, e.g.
\begin{equation}
c_i^{ren} = c_i + {\gamma_i  \Lambda^2 \over 16 \pi^2}
\end{equation}
for some specific coefficient $c_i$. (Here $\gamma_i$ is a number to be
calculated in the renormalization process.) All phenomenology can be
expressed in terms of the renormalized parameters and the strong
$\Lambda^2$ depencence of this example would have vanished.
When the meson masses are small, we will Taylor expand the loop
integrals, renormalize the chiral parameters and recover exactly the
usual results.

For realistic phenomenology, however, we need to use the physical
values of the meson masses. The kaon and eta masses are in reality not
small compared to the separation scale. They are also not so large
that all of their effects can reliably treated as short-distance and
hence be built into the parameters of the effective Lagrangian. We do
need to treat them as dynamical degrees of freedom and include at
least their long distance effects. When we use a cutoff
regularization, with a cutoff close to the separation scale, the loop
integrals will be the nonlinear functions of the mass, as described in
the previous section. When these are evaluated at the physical meson
masses, this will generate effects that are equivalent to higher
orders in the chiral expansion. Thus this form of regularization can
be viewed as a partial resummation of the chiral series. If we
continue to treat the problem in full generality, we will still need
to include chiral parameters in the effective Lagrangian which will
allow us to continue to be fully model-independent. In each of 
the sections
that follow, we will explore the phenomenology at physical values of
the meson masses.

The cutoff $\Lambda$ should not be taken so low in energy that it
removes any truly long distance physics. Also, while it can in principle be
taken much larger than the separation scale, this will lead to the
inclusion of spurious short distance physics which can upset the
convergence of the expansion. It is ideal to take the cutoff slightly
above the separation scale so that all of the long distance physics,
but little of the short distance physics, is included.

This procedure is not a model. Indeed its purpose is to remove the
model-dependent short distance portions of loops. However, it appears
to do so at the cost of introducing a new parameter, the cutoff
$\Lambda$, plus the dependence of the choice of cutoff function. If
the form of this function or the value of $\Lambda$ played a major
role in the phenomenology, this would be a serious drawback for this
approach. However, renormalization theory leads us to expect that the
dependence on the cutoff should be quite mild in phenomenological
applications. This is because the cutoff (and the functional
dependence) can be absorbed in the renormalization of the chiral
parameters. If one worked to all orders, all dependence would
disappear. If one is working to a given finite order, the residual
dependence is expected to occur only at the next order beyond 
that at which one is working at. Since it appears from the 
above analysis 
that the cutoff integrals are rather slowly varying functions of
the mass, we expect that working to an order where one includes the
first SU(3) breaking parameters should be sufficient to minimize the
cutoff dependence to an acceptable value.

Another issue that we should address here is the nature of the energy
expansion in such a procedure. When using a regularization scheme
which does not contain any dimensionful parameters, there is a
particularly simple power counting procedure which determines the
order of contributions of loop diagrams. If the regularization scheme
does involve a mass parameter, this counting will not directly apply. 
We
will see this explicitly below as the loop process will renormalize
chiral parameters at different orders in the energy expansion. As one
goes to the next order in loops, one will have to perform this
renormalization again order by order. This, however, is not a
fundamental problem. As we show, for small values of the meson masses 
we obtain exactly the same results as in other regularization schemes.
Therefore, we can use the small mass limit to set up the chiral
expansion and determine the order of the loops that one should
include. Subsequently taking the masses to their physical values will
accomplish the partial resummation of effects described above.
However, the procedure in terms of which loops to include need not be
changed.  

There is one special feature involved in doing chiral perturbation
theory with a cutoff instead of dimensional regularization. This
involves an occasional change in the Feynman rules due to the presence
of derivative couplings. The analysis of this aspect goes back to
a classic
paper on the subject, \cite{Gerstein}. 
Recall that in the canonical construction of
a field theory, one forms the canonical momenta conjugate to the field
variables via
\begin{equation}
\pi(x) = {\delta {\cal L} \over \delta \partial_0 \phi(x)} \ \ .
\end{equation}
When the interaction piece of the Lagrangian involves time
derivatives, the canonical momenta will also carry portions of the
interaction so that in forming the Hamiltonian, the interaction Hamiltonian
will no longer be simply the negative of the interaction Lagrangian.
Since perturbation theory and the Feynman rules are formulated from
the interaction Hamiltonian, the canonical formalism will involve some
modified (and non-covariant) vertices. At the same time, the presence
of time derivatives in interactions will act on the time ordering in
propagators to produce further non-covariant contributions to loop
processes\cite{Gerstein}. 
These modifications do not always cancel but can leave a
residual interaction. While one can simply calculate this using the
straightforward but clumsy canonical formalism, 
the authors of Ref. \cite{Gerstein}
show that one can use the naive rules if one adds a specific contact
interaction proportional to $\delta^4(0)$ to the Feynman rules of the
mesonic part of the theory. When using dimensional regularization, one
of the peculiarities is that the regularized value of $\delta^4(0)$ is
equal to zero. Therefore the contact interaction vanishes and we may
proceed using the naive Feynman rules when calculating dimensionally.
However, with a momentum-space cutoff, one has $\delta^4(0) \sim
\Lambda^4$ and one gets a nontrivial modification quartic in the
cutoff. This influences the purely mesonic sector of the theory. We
have verified, however, that the baryonic processes that we consider are not
modified by this feature at the order that we are working.

We now explore several specific cases of the physics of
loop processes in baryon chiral perturbation theory. Our procedure in
each case is the same. We take the known results of a standard
analysis of the one-loop amplitudes and re-express it in terms of the
Feynman integrals that we have analysed. We then show how the
renormalization procedure is accomplished with a cutoff, absorbing
the leading cutoff dependence into renormalized parameters. In each
case this reproduces the standard analysis for small values of the
mesonic masses. Then we turn to the realistic case of the physical
meson masses and a finite cutoff. In this situation, the presence of
the cutoff only permits the long distance loop effects, and this leads
to the much more moderate effect of loops compared to the results quoted
in the introduction.

\section{Baryon Masses}

In this section we return to the physics of chiral loops, as illustrated
in the analysis of baryon masses, and deal with specific numerical results.
This has already been discussed in
Ref \cite{DH}, but it is pedagogically useful to revisit the analysis in the
present
context. This will clearly illustrate the renormalization program and the
isolation
of long-distance loop effects.

To lowest and next leading order
in the derivative expansion the effective Lagrangian which descibes
the interactions of baryons can be written, in the heavy baryon formalism, as
\begin{eqnarray}
{\cal L}_MB&=&{\rm Tr}\bar{B}iv\cdot DB +
d_A{\rm Tr}\bar{B}S^\mu\{u_\mu,B\} +f_A{\rm
Tr}\bar{B}S^\mu[u_\mu,B]\nonumber\\
&+&d_m{\rm Tr}\bar{B}\{\chi_+,B\}
+f_m{\rm Tr}\bar{B}[\chi_+,B]+b_0{\rm Tr}\bar{B}B{\rm Tr}\chi_+\ldots
\end{eqnarray}\label{eq:uu}
where $\chi_+$ is given in terms of the quark mass matrix $m$
via $\chi_+=2B_0m$,
\begin{equation}
D_\mu=\partial_\mu+{1\over 2}[u^\dagger,\partial_\mu u]
\end{equation}
is the covariant derivative, and
\begin{equation}
S_\mu={i\over 2}\gamma_5\sigma_{\mu\nu}v^\nu
\end{equation}
is the Pauli-Lubanski spin vector.  The nonlinear mesonic
chiral constructs $u,u_\mu$ are given by
\begin{equation}
U=u^2=\exp({i\over F_\pi}\sum_j\lambda_j\phi_j),\qquad
u_\mu=iu^\dagger\partial_\mu U u^\dagger
\end{equation}
Here $M_0,f_m,d_m,b_0$ are free parameters in
terms of which the tree level contribution to the baryon masses can be
written as given above in Eq. \ref{eq:cc} with
\begin{equation}
\hat{M}_0=M_0-2(2m_K^2+m_\pi^2)b_0
\end{equation}

If we continue the analysis to higher order we include the effects
of quark loops and of the higher order terms in a general Lagrangian.
In an expansion in quark mass we
have the schematic form
\begin{equation}
M_B=M_0+\sum_qa_qm_q+\sum_qb_qm_q^{3\over 2}+\sum_qc_qm_q^2+\ldots
\end{equation}
Here, the terms linear in the quark mass are those parametrized in Eq.
\ref{eq:cc},
where we recall that $m_P^2 \sim m_q$. The next term in the expansion is
nonanalytic in
the quark mass and comes uniquely from
loop diagrams. Finally the terms at order $m_q^2$ come from yet higher
order effects which we will not explicitly consider here.

The one loop chiral corrections are well known and involve the integral
given in Eq. \ref{eq:pp} of the previous section. In dimensional
regularization
this yields
terms in $m_P^3$ and can be represented as
\begin{equation}
\delta M_i=-{1\over 24\pi F_\pi^2}\sum_j \kappa_i^jm_j^3\label{eq:dd}
\end{equation}
with
\begin{eqnarray}
\kappa_N^\pi&=&{9\over
4}(d_A+f_A)^2,\quad \kappa_N^K={1\over
2}(5d_A^2-6f_Ad_A+9f_A^2),\quad \kappa_N^\eta={1\over 4}(d_A-3f_A)^2\nonumber\\
\kappa_\Sigma^\pi&=&(d_A^2+6f_A^2),\quad
\kappa_\Sigma^K=3(d_A^2+f_A^2),\quad
\kappa_\Sigma^\eta=d_A^2\nonumber\\
\kappa_\Lambda^\pi&=&3d_A^2,\quad\kappa_\Lambda^K=d_A^2+9f_A^2,\quad
\kappa_\Lambda^\eta=d_A^2\nonumber\\
\kappa_\Xi^\pi&=&{9\over
4}(d_A-f_A)^2,\quad\kappa_\Xi^K={1\over
2}(5d_A^2+6d_Af_A+9f_A^2),\quad
\kappa_\Xi^\eta={1\over 4}(d_A+3f_A)^2\nonumber\\
\quad
\end{eqnarray}
This produces the large mass shifts quoted in Eq. 3.
The violation of the Gell-Mann-Okubo relation is given then by
\begin{equation}
{1\over 4}[3M_\Lambda+M_\Sigma-2M_N-2M_\Xi]={d_A^2-3f_A^2\over
96\pi F_\pi^2}[4m_K^3-3m_\eta^3-m_\pi^3]
\end{equation}
The deviation from the Gell-Mann-Okubo relation due to loops is found to
be quite small, primarily due to the (accidental) feature that
$d_A^2-3f_A^2\approx 0.02 <<1$.

We now turn to an exploration of the analysis using a cutoff regulariation.
The first task is to see how the renormalization program works, 
in order that
we obtain exactly the same result in the limit of small masses.
The diagrams involved are the same as in the previous analysis, but
we utilize the cutoff form for the Feynman integral. This is simply done
by replacing $m_P^3$ in Eq. \ref{eq:dd} by the function $I_\Lambda (m_P^2)$,
expanded as in Eq. \ref{eq:ee}.  The one loop contribution to the mass
then has the schematic form
\begin{equation}
\delta M_i=-{1\over 24\pi F_\pi^2}\sum_j({1\over 2}\Lambda^3-{1\over
2}\Lambda m_j^2+m_j^3+\ldots)
\end{equation}
Obviously the term in $m_j^3 $ is
identical to that arising in conventional dimensional regularization,
but more interesting are the contributions proportional to
$\Lambda^3$ and to $\Lambda m_P^2$.  The piece cubic in $\Lambda$ has the form
\begin{equation}
\delta M_i^{\Lambda^3}=-{\Lambda^2\over 48\pi
F_\pi^2}\sum_j\kappa_i^j
\end{equation}
and is independent of baryon type---it may be absorbed into a
renormalization of $M_0$---
\begin{equation}
M_0^r=M_0-(5d_A^2+9f_A^2){\Lambda^3\over 48\pi F_\pi^2}
\end{equation}
On the other hand the terms linear in $\Lambda$
\begin{equation}
\delta M_i^{\Lambda}={\Lambda\over 48\pi F_\pi^2}\sum_j\kappa_i^jm_j^2
\end{equation}
must be able to be absorbed into renormalizations of the coefficients
involving $m_q$, and indeed this is found to be the case---one
verifies that
\begin{eqnarray}
d_m^r&=&=d_m-{3f_A^2-d_A^2\over 128\pi F_\pi^2}\Lambda\nonumber\\
f_m^r&=&f_m-{5d_Af_A\over 192\pi F_\pi^2}\Lambda\nonumber\\
b_0^r&=&b_0-{13d_A^2+9f_A^2\over 576\pi F_\pi^2}\Lambda
\end{eqnarray}
That such renormalization can occur involves a highly constrained set
of conditions and the fact that they are satisfied is a significant
verification of the chiral invariance of the cutoff procedure.  Of
course, once one has defined renormalized coefficients, since they are
merely phenomenological parameters which must be determined empirically,
the procedure is identical to the results of the usual dimensionally
regularized technique when the masses are smaller than the cutoff.

Having convinced ourselves of the chiral invariance of the cutoff
procedure to the order we are working,
we can now apply it to the case where masses are their physical values
and the cutoff is taken
to be phenomenologically relevant----{\it i.e.},
$\Lambda\geq 1/<r_B>\sim 300-600$
MeV.  However, we first remove the asymptotic mass-independent component of
the function $I(m)$ by defining
\begin{equation}
\tilde{I}(m)=I(m)-{1\over 2}\Lambda^3
\end{equation}
since these effects can be absorbed into $M_0$ and give misleading
indications about the size of the nonanalytic effects in the large
cutoff limit.  The size of the long distance nonanalytic contributions to the
baryon masses is then given by
\begin{equation}
\delta M_i=-{1\over 24\pi F_\pi^2}\sum_j\kappa_i^j\tilde{I}(m_j)
\end{equation}
and the corresponding numerical results are given in Table 1.  A
careful look at these findings reveals that the quantitative results
are in agreement with our qualitative expectations---for a reasonable
value of the cutoff parameter $\Lambda$, the overall size of the nonanalytic
corrections
is much smaller that that found in the dimensionally regularized case
since the short distance contribution from kaon, eta loops is much reduced.
 There is
no longer any in principle problem with the convergence of the chiral
expansion and the ``mystery'' of why the lowest order fit linear in
$m_q$ works so well is resolved.  Of course, one still must include
the model-dependent contribution from short distance effects, but
there no longer exists a problem from the calculable and
model-independent long distance component.

\begin{table}
\begin{center}
\begin{tabular}{c|c|c|c|c|c}
  & dim. & $\Lambda=300$&$\Lambda=400$&$\Lambda=500$&$\Lambda=600$\\
\hline
$N$&-0.31&0.02&0.03&0.05&0.07\\
$\Sigma$&-0.62&0.03&0.05&0.08&0.12\\
$\Lambda$&-0.68&0.03&0.06&0.09&0.13\\
$\Xi$&-1.03&0.04&0.08&0.12&0.17
\end{tabular}
\caption{Given (in GeV) are the nonanalytic contributions to baryon
masses in dimensional regularization and for various values of the
cutoff parameter $\Lambda$ in MeV.}
\end{center}
\end{table}
A good fit to the baryon masses can be accomplished 
for any value of the cutoff
in the range that we consider. For example, with $\Lambda = 400$MeV, we have
the
masses described by
\begin{eqnarray}
M_N  & = & 1.143 - 0.237 + 0.034 = 0.940     \nonumber          \\
M_\Sigma & = & 1.143 - 0.005 + 0.053 =1.191  \nonumber         \\
M_\Lambda & = &  1.143 - 0.086 + 0.057 = 1.114  \nonumber       \\
M_\Xi & = &     1.143 + 0.106 + 0.077 = 1.326    \label{eq:vv}
\end{eqnarray}
where all numbers are given in GeV.
In Eq.~\ref{eq:vv}, $\hat{M}_0$ is the first term,
the second term comes from the leading tree level SU(3) breaking
due to quark masses parameterized as in Eq. \ref{eq:cc} and the last term
from the residual loop effects. The tree level terms contribute 343 MeV
to the $\Xi$-N mass splitting, while the loop effects contribute only
43 MeV.  The chiral expansion is well-behaved---loops do not upset the
basic pattern at lowest order and the approximate
SU(3) invariance is manifest. In order to disentangle $M_0$ and $b_0$, one
has also to take, {\it e.g}, the $\pi N$ $\sigma$--term into account \cite{B}.

If we had used a different value of the cutoff in the regularization,
the specific contributions would have been different, yet the final
answers change by less that 1 MeV for $\Lambda$ from 300 MeV to 600
MeV. This is a demonstration of the cutoff independence of this
procedure. (Our previous discussion suggested that we should have
found a cutoff dependence equivalent to neglected higher order terms,
which in this case would have been of order 5 MeV. In practice we
found less dependence than that.) We have also verified that we obtain
identical results for another form of the cutoff function\cite{DH}. 

Having seen how the cutoff procedure can be successfully applied in
the case of the baryon masses, we can now move on the the remaining
applications -- axial coupling, nonleptonic hyperon decay, and magnetic
moments -- to show how a chirally consistent picture emerges therein.

\section{Axial Currents}

The baryon axial couplings are parameterized in terms of the same $f_A,d_A$
coofficients
which appear in the Hamiltonian of Eq \ref{eq:uu}. Defining the lowest order
contribution
using the notation $g_A(\bar{i}j)=\alpha_{ij}$, we have
\begin{eqnarray}
\alpha_{pn}&=&f_A+d_A\nonumber\\
\alpha_{\Lambda\Sigma^-}&=&{2\over \sqrt{6}}d_A\nonumber\\
\alpha_{p\Lambda}&=&-{1\over \sqrt{6}}(d_A+3f_A)\nonumber\\
\alpha_{\Lambda\Xi^-}&=&-{1\over \sqrt{6}}(d_A-3f_A)\nonumber\\
\alpha_{n\Sigma^-}&=&d_A-f_A\nonumber\\
\alpha_{\Sigma^0\Xi^-}&=&{1\over \sqrt{2}}\alpha_{\Sigma^+\Xi^0}
={1\over \sqrt{2}}(d_A+f_A)
\end{eqnarray}
It is these forms which are used in SU(3) fits to hyperon beta decay.

The leading
nonanalytic corrrections from loops
are ${\cal O}(m_P^2\ln m_P^2)$ and were first
calculated by Bijnens, Sonoda, and Wise\cite{BSW}.  They have the form
\begin{equation}
g_A(\bar{i}j)=\sqrt{Z_iZ_j}[\alpha_{ij}+{1\over 16\pi^2
F_\pi^2}\sum_k \beta_{ij}^k m_k^2\ln {m_k^2\over \mu^2}]
\end{equation}
with
\begin{eqnarray}
\beta_{pn}^\pi&=&{1\over
4}(d_A^3+f_A^3+3d_A^2f_A+3f_A^2d_A)-(d_A+f_A),\nonumber\\
\beta_{pn}^K&=&{1\over
3}d_A^3-{1\over 3}f_Ad_A^2+d_Af_A^2-f_A^3-{1\over
2}(d_A+f_A),\nonumber\\
 \beta_{pn}^\eta&=&-{1\over 12}d_A^3+{5\over 12}f_Ad_A^2-{1\over
4}d_Af_A^2-{3\over 4}f_A^3\nonumber\\
\beta_{p\Lambda}^\pi&=&{1\over \sqrt{6}}(-{3\over 2}d_A^3+{3\over
2}d_Af_A^2+{3\over 8}(d_A+3f_A)),\nonumber\\
\beta_{p\Lambda}^K&=&{1\over \sqrt{6}}({5\over
6}d_A^3-{5\over 2}d_A^2f_A-{3\over 2}f_A^2d_A+{9\over 2}f_A^3)+{3\over
4}(d_A+3f_A)),\nonumber\\
\beta_{p\Lambda}^\eta&=&{1\over \sqrt{6}}({1\over
6}d_A^3-{3\over 2}d_Af_A^2+{3\over 8}(d_A+3f_A))\nonumber\\
\beta_{\Lambda\Sigma^-}^\pi&=&{1\over \sqrt{6}}(-{2\over
3}d_A^3+2d_Af_A^2-2d_A),\nonumber\\
\beta_{\Lambda\Sigma^-}^K&=&{1\over
\sqrt{6}}(d_A^3-d_Af_A^2-d_A),\nonumber\\
\beta_{\Lambda\Sigma^-}^\eta&=&{1\over \sqrt{6}}({2\over
3}d_A^3)\nonumber\\
\beta_{n\Sigma^-}^\pi&=&{1\over 6}d_A^3-{1\over 3}d_A^2f_A+{2\over
3}d_Af_A^2+f_A^3-{3\over
8}(d_A-f_A)),\nonumber\\
\beta_{n\Sigma^-}^K&=&{1\over 2}f_A^3+{1\over 2}d_Af_A^2+{1\over
6}d_A^2f_A+{1\over 6}d_A^3-{3\over
4}(d_A-f_A)),\nonumber\\
\beta_{n\Sigma^-}^\eta&=&{1\over 2}d_Af_A^2-{2\over
3}d_A^2f_A+{1\over 6}d_A^3-{3\over 8}(d_A-f_A))\nonumber\\
\beta_{\Lambda\Xi^-}^\pi&=&{1\over \sqrt{6}}(-{3\over 2}d_A^3
+{3\over 2}f_A^2d_A+{3\over
8}(d_A-3f_A)),\nonumber\\
\beta_{\Lambda\Xi^-}^K&=&{1\over \sqrt{6}}({5\over
6}d_A^3+{5\over 2}d_A^2f_A-{3\over 2}d_Af_A^2-{9\over 2}f_A^3+{3\over
4}(d_A-3f_A)),\nonumber\\
\beta_{\Lambda\Xi^-}^\eta&=&{1\over
\sqrt{6}}({1\over 6}d_A^3-{3\over 2}d_Af_A^2+{3\over 8}(d_A-3f_A))\nonumber\\
\beta_{\Sigma^0\Xi^-}^\pi&=&{1\over \sqrt{2}}(-f_A^3+{1\over
3}f_Ad_A^2+{1\over 2}f_A^2d_A+{1\over 6}d_A^3-{3\over
8}(d_A+f_A)),\nonumber\\
\beta_{\Sigma^0\Xi^-}^K&=&{1\over \sqrt{2}}({1\over
6}d_A^3-{1\over 6}f_Ad_A^2+{1\over 2}f_A^2d_A-{1\over 2}f_A^3-{3\over
4}(d_A+f_A)),\nonumber\\
\beta_{\Sigma^0\Xi^-}^\eta&=&{1\over \sqrt{2}}({1\over 6}d_A^3+{2\over
3}d_A^2f_A+{1\over 2}d_Af_A^2-{3\over 8}(d_A+f_A))
\end{eqnarray}
Here $Z_i$ are the wavefunction renormalization factors, whose leading
nonanalytic form is
\begin{equation}
Z_i=1-{1\over 16\pi^2 F_\pi^2}\sum_j \kappa_i^j m_j^2\ln {m_j^2\over \mu^2}
\end{equation}
with $\kappa_i^j$ given in Eq. 39.
These forms generate the corrections discussed in the introduction.

When we apply the cutoff formalism we first note that
all of the nonanalytic behavior
of the form $m^2 \ln m^2$ comes uniquely
from the integral that we labeled $J(m)$ in
Section 3. This means that all that we need to do
in order to convert the analysis above to
our formalism is to replace $m_P^2 \ln m_P^2$ by $J(m_P)$ everywhere throughout
these
formulas.
We may again check
the chiral consistency of the renormalization program by verifying that
the contribution quadratic in $\Lambda$---
\begin{equation}
\delta g_A^{\Lambda^2}(\bar{i}j)={\Lambda^2\over 16\pi^2F_\pi^2}
\sum_k[\beta_{ij}^k-{1\over 2}\alpha_{ij}(\lambda_i^k+\lambda_j^k)]
\end{equation}
can be absorbed into renormalizations of the lowest order axial
couplings $d_A,f_A$ via
\begin{eqnarray}
d_A^r&=&d_A-{3\over 2}d_A(3d_A^2+5f_A^2+1){\Lambda^2\over 16\pi^2
F_\pi^2}\nonumber\\
f_A^r&=&f_A-{1\over 6}f_A(25d_A^2+63f_A^2+9){\Lambda^2\over 16\pi^2 F_\pi^2}
\end{eqnarray}
Since such coefficients are determined empirically the analysis with
small meson masses becomes
identical to that of the dimensionally regularized case.

In the case of a physically realistic cutoff---$\Lambda\sim 300-600$ MeV---
and the physical meson masses, we
have
\begin{equation}
\delta g_A(\bar{i}j)={1\over 16\pi^2
F_\pi^2}\sum_k[\beta_{ij}^k-{1\over
2}\alpha_{ij}(\lambda_i^k+\lambda_j^k)]\tilde{J}(m_k^2)
\end{equation}
where we have again removed the asymptotic mass-independent component of
the function $J(M^2)$ via
\begin{equation}
\tilde{J}(m^2)=J(m^2)-\Lambda^2
\end{equation}
The numerical results using typical values of the cutoff are
compared with those from dimensional regularization in Table 2 and
again reflect the feature that
the SU(3) chiral expansion is now under control at least as far as
long distance effects are concerned---the ``mystery'' of the
correctness of the simple SU(3) fit {\it without} chiral corrections
is resolved. A complete discussion of axial-vector current matrix
elements can be found in \cite{B1}.

\begin{table}
\begin{center}
\begin{tabular}{c|c|c|c|c|c}
  &dim.&$\Lambda$=300&$\Lambda$=400&$\Lambda$=500&$\Lambda$=600\\
\hline
$g_A(\bar{p}n)$&1.72&0.37&0.53&0.69&0.84\\
$g_A(\bar{p}\Lambda)$&-1.78&-0.34&-0.51&-0.67&-0.84\\
$g_A(\bar{\Lambda}\Sigma^-)$&1.17&0.23&0.34&0.45&0.56\\
$g_A(\bar{n}\Sigma^-)$&0.36&0.07&0.10&0.14&0.17\\
$g_A(\bar{\Lambda}\Xi^-)$&0.83&0.15&0.23&0.31&0.39\\
$g_A(\bar{\Sigma^0}\Xi^-)$&2.46&0.45&0.68&0.91&1.15
\end{tabular}
\caption{Given are the nonanalytic contribtions to $g_A$ for various
transitions in dimensional regularization and for various values of
the cutoff parameter $\Lambda$ in MeV.}
\end{center}
\end{table}

\section{S-wave hyperon decay}
Chiral invariance relates the S-wave nonleptonic decay amplitudes to the
baryon-to-baryon matrix elements of the weak Hamiltonian. For the dominant
octet Hamiltonian this can be paramterized in terms of two SU(3)
coefficients$f_w,d_w$:
\begin{equation}
A(\Upsilon_i^j)=\zeta(\Upsilon^i_j)
\end{equation}
where
\begin{eqnarray}
\zeta(\Lambda^0_0)&=&-{1\over \sqrt{2}}\zeta(\Lambda^0_-)=-{1\over
2\sqrt{3}}(d_w+3f_w)\nonumber\\
\zeta(\Sigma^+_0)&=&-{1\over \sqrt{2}}\zeta(\Sigma^-_-)={1\over
\sqrt{2}}(d_w-f_w)\nonumber\\
\zeta(\Sigma^+_+)&=&0\nonumber\\
\zeta(\Xi^0_0)&=&-{1\over \sqrt{2}}\zeta(\Xi^-_-)=-{1\over 2\sqrt{3}}(d_w-3f_w)
\end{eqnarray}
This yields a good fit to the data, including the chiral SU(3) results
given in Eq. 5.

Proceeding to one-loop order, the
leading nonanalytic corrections are dependent upon $m_P^2\ln
m_P^2$ and have the form
\begin{equation}
A(\Upsilon^i_j)=\sqrt{Z_iZ_j}[\zeta(\Upsilon^i_j)+
{1\over 16\pi^2 F_\pi^2}\sum_k \rho(\Upsilon^i_j)^km_k^2\ln
{m_k^2\over \mu^2}]
\end{equation}
with
\begin{eqnarray}
\rho(\Lambda^0_0)^\pi&=&-{1\over 2\sqrt{3}}d_w({7\over 24}-{9\over
2}d_A^2-{9\over 2}d_Af_A)
-{1\over 2\sqrt{3}}f_w({7\over 8}+{9\over 2}d_A^2+{9\over 2}d_Af_A)\nonumber\\
\rho(\Lambda_0^0)^K&=&-{1\over
2\sqrt{3}}d_w(-{5\over 12}+{5\over 2}d_A^2-9f_Ad_A+{9\over 2}f_A^2)\nonumber\\
&-&{1\over
2\sqrt{3}}f_w(-{5\over 4}
+{3\over 2}d_A^2-9f_Ad_A+{27\over 2}f_A^2)\nonumber\\
\rho(\Lambda^0_0)^\eta&=&-{1\over 2\sqrt{3}}d_w(-{3\over 8}+{1\over
2}d_A^2-{3\over 2}d_Af_A)
-{1\over 2\sqrt{3}}f_w(-{9\over 8}+{3\over 2}d_A^2-{9\over 2}f_Ad_A)\nonumber\\
\rho(\Xi^0_0)^\pi&=&-{1\over 2\sqrt{3}}d_w({7\over 24}-{9\over 2}d_A^2+
{9\over 2}d_Af_A)
+{1\over 2\sqrt{3}}f_w({7\over 8}+{9\over 2}d_A^2-{9\over 2}d_Af_A)\nonumber\\
\rho(\Xi^0_0)^K&=&-{1\over 2\sqrt{3}}d_w(-{5\over 12}+{5\over
2}d_A^2+9d_Af_A+{9\over 2}f_A^2)\nonumber\\
&+&{1\over 2\sqrt{3}}f_w(-{5\over 4}+{3\over
2}d_A^2+9d_Af_A+{27\over 2}f_A^2)\nonumber\\
\rho(\Xi^0_0)^\eta&=&-{1\over 2\sqrt{3}}d_w(-{3\over 8}+{1\over 2}d_A^2+{3\over
2}d_Af_A)
+{1\over 2\sqrt{3}}f_w(-{9\over 8}+{3\over 2}d_A^2+{9\over 2}d_Af_A)\nonumber\\
\rho(\Sigma^+_0)^\pi&=&\sqrt{1\over 2}d_w({7\over 24}+3f_A^2
+{5\over2}d_Af_A-{1\over 2}d_A^2)\nonumber\\
&-&{1\over \sqrt{2}}f_w({7\over 24}+3f_A^2+{9\over 2}d_Af_A+{3\over
2}d_A^2)\nonumber\\
\rho(\Sigma^+_0)^K&=&{1\over \sqrt{2}}d_w(-{5\over 12}-{1\over
2}d_A^2+d_Af_A+{3\over 2}f_A^2)\nonumber\\
&-&{1\over \sqrt{2}}
f_w(-{5\over 12}+{3\over 2}d_A^2+3d_Af_A+{3\over 2}f_A^2)\nonumber\\
\rho(\Sigma^+_0)^\eta&=&{1\over \sqrt{2}}d_w(-{3\over 8}-{1\over
2}d_A^2+{3\over 2}d_Af_A)
-{1\over \sqrt{2}}f_w(-{3\over 8}-{1\over 2}d_A^2+{3\over 2}d_Af_A)\nonumber\\
\rho(\Sigma^+_+)^\pi&=&\rho(\Sigma^+_+)^K=\rho(\Sigma^+_+)^\eta=0
\end{eqnarray}
The correction to the Lee-Sugawara relation is found
\begin{eqnarray}
&&\sqrt{3}A(\Sigma^+_0)-2A(\Xi^-_-)-A(\Lambda^0_-)
=-\sqrt{2\over 3}{1\over 16\pi F_\pi^2}\nonumber\\
&&\times[m_K^2\ln m_K^2(d_w({9\over 2}d_A^2+3d_Af_A+{9\over 2}f_A^2)
+f_w({3\over 2}d_A^2-9d_Af_A-{9\over 2}f_A^2))\nonumber\\
&&+m_\eta^2\ln m_\eta^2(d_w({3\over 2}d_A^2-{3\over 2}d_Af_A)
+f_w(-{3\over 2}d_A^2-{9\over 2}d_Af_A))\nonumber\\
&&+m_\pi^2\ln m_\pi^2(d_w(-6d_A^2-{3\over 2}d_Af_A-{9\over 2}f_A^2)
+f_w({27\over 2}d_Af_A+{9\over 2}f_A^2))]\nonumber\\
&&\approx -6.4\times 10^{-7}
\end{eqnarray}
When analysed using the physical values of the masses, we uncover the problems
described in the introduction.

A very similar analysis obtains as was describe in the situation
for the axial currents. In the cutoff formalism the
nonanalytic pieces proportional to $m_P^2\ln m_P^2$
are simply replaced by the function $J(m_P^2)$.  Again, the chiral
consistency of of the renormalization program can be
verified by noting that for small meson masses the
component quadratic in $\Lambda$---
\begin{equation}
\delta A(\Upsilon^i_j)={\Lambda^2\over 16\pi^2
F_\pi^2}[\rho(\Upsilon^i_j)^k-{1\over 2}\zeta(\Upsilon^i_j)
(\lambda_i^k+\lambda_j^k)]
\end{equation}
can be absorbed into renormalized values of the lowest order couplings
$f_w,d_w$
via
\begin{eqnarray}
d_w^r&=&d_w-{1\over
2}[d_w(1+13d_A^2+9f_A^2)+18f_wd_Af_A]{\Lambda^2\over 16\pi^2
F_\pi^2}\nonumber\\
f_w^r&=&f_w-{1\over
2}[f_w(1+5d_A^2+9f_A^2)+10d_wd_Af_A]{\Lambda^2\over 16\pi^2 F_\pi^2}
\end{eqnarray}
Once this renormalization is accomplished, we exactly
recover the usual chiral analysis.

In the case of a physically realistic masses, we again use the
same mass-independent renormalization to define the residual integral
${\tilde{J}}(m)$. The shift in s-wave
amplitudes is then given by
\begin{equation}
\delta A(\Upsilon^i_j)=\sum_k[\rho(\Upsilon^i_j)^k-{1\over
2}\zeta(\Upsilon^i_j)(\lambda_i^k+\lambda_j^k)]{\tilde{J}(m_k^2)\over 16\pi^2
F_\pi^2}
\end{equation}
The numerical results are compared with those of dimensional
regularization in Table 3 and it is clear that once again the results
are dominated by the lowest order SU(3) forms---there no longer exist
large chiral corrections.

\begin{table}
\begin{center}
\begin{tabular}{c|c|c|c|c|c}
 &dim.&$\Lambda$=300&$\Lambda$=400&$\Lambda$=500&$\Lambda$=600\\
\hline
$A(\Lambda^0_0)$&-3.57&-0.62&-0.95&-1.30&-1.65\\
$A(\Xi^0_0)$&1.96&0.36&0.54&0.73&0.92\\
$A(\Sigma^+_0)$&-1.57&-0.26&-0.41&-0.56&-0.72
\end{tabular}
\caption{Given are the nonanalytic contributions to s-wave
semileptonic hyperon decay amplitudes in dimensional regularization
and for various values of the cutoff parameter $\Lambda$ in MeV.}
\end{center}
\end{table}

\section{Magnetic moments}

The final case considered here is that of magnetic moments. The lowest order
parameterization is given in Eq. \ref{eq:yy}. The leading nonanalytic
chiral corrections are linear in $m_P$ and were first calculated by
Caldi and Pagels.  They have the form
\begin{equation}
\delta \mu_i= {M_0\over 8\pi F_\pi^2}\sum_j \sigma_i^j m_j
\end{equation}
with
\begin{eqnarray}
\sigma_p^\pi&=&-(f_A+d_A)^2,\quad\sigma_p^K=-{2\over
3}(d_A^2+3f_A^2)\nonumber\\
\sigma_n^\pi&=&(d_A+f_A)^2,\quad\sigma_n^K=-(d_A-f_A)^2\nonumber\\
\sigma_\Lambda^\pi&=&0,\quad\sigma_\Lambda^K=2f_Ad_A\nonumber\\
\sigma_{\Sigma^+}^\pi&=&-{2\over 3}(d_A^2+3f_A^2),\quad\sigma_{\Sigma^+}^K
=-(d_A+f_A)^2\nonumber\\
\sigma_{\Sigma_0}^\pi&=&0,\quad\sigma_{\Sigma^0}^K=-2d_Af_A\nonumber\\
\sigma_{\Sigma^-}^\pi&=&{2\over
3}(d_A^2+3f_A^2),\quad\sigma_{\Sigma^-}^K=(d_A-f_A)^2\nonumber\\
\sigma_{\Lambda\Sigma}^\pi&=&-{4\over \sqrt{3}}d_Af_A,\quad
\sigma_{\Lambda\Sigma}^K=-{2\over \sqrt{3}}d_Af_A\nonumber\\
\sigma_{\Xi^-}^\pi&=&(d_A-f_A)^2,\quad\sigma_{\Xi^-}^K
={2\over 3}(d_A^2+3f_A^2)\nonumber\\
\sigma_{\Xi^0}^\pi&=&-(d_A-f_A)^2,\quad\sigma_{\Xi^0}^K=(d_A+f_A)^2
\end{eqnarray}

In this analysis, all Feynman integrals are given by the linear form
called $K(m)$ in Section 3. The general result appropriate for a cutoff
regularization is obtained by replacing the nonanalytic dependence $m_P$ by
$K(m_P)$. We can then verify that the leading term in $\Lambda$ can be
absorbed into the renormalization of the chiral parameters, leading
to an identical analysis for small values of $m_P$. In this case, examination
of the
term in the magnetic moment shift linear in $\Lambda$---
\begin{equation}
\delta\mu_i^\Lambda=-{M_0\Lambda\over 24\pi F_\pi^2}\sum_j\sigma_i^j
\end{equation}
shows that it be absorbed into renormalizations of the lowest order
parameters $f_\mu,d_\mu$ via
\begin{eqnarray}
d_\mu^r&=&d_\mu+{M_0\Lambda\over 4\pi F_\pi^2}d_Af_A\nonumber\\
f_\mu^r&=&f_\mu+{M_0\Lambda\over 24\pi F_\pi^2}({5\over 3}d_A^2+3f_A^2)
\end{eqnarray}
Since $f_\mu,d_\mu$ are determined empirically, the analysis is then
identical to that of the dimensionally regularized case.

On the other hand with the use of a physically realistic cutoff and meson
masses, the magnetic
moment shifts cam be obtained by using the mass independent renormalization
given by
\begin{equation}
\tilde{K}(m)=K(m)+{1\over 3}\Lambda
\end{equation}
The shifts in the magnetic moments
are given by
\begin{equation}
\delta\mu_i={M_0\over 8\pi F_\pi^2}\sum_j\sigma_i^j\tilde{K}(m_j)
\end{equation}
The numerical results for this form for reasonable values of the
cutoff are compared with those from dimensional regularization in
Table 4.  Again the chiral corrections are no longer out of control.

\begin{table}
\begin{center}
\begin{tabular}{c|c|c|c|c|c}
 &dim.&$\Lambda$=300&$\Lambda$=400&$\Lambda$=500&$\Lambda$=600\\
\hline
$\mu_p$&0.76&0.22&0.27&0.31&0.34\\
$\mu_n$&-0.22&-0.12&-0.14&-0.15&-0.16\\
$\mu_\Lambda$&-0.43&-0.08&-0.11&-0.13&-0.15\\
$\mu_{\Sigma^+}$&1.05&0.24&0.30&0.36&0.40\\
$\mu_{\Sigma^0}$&0.44&0.08&0.11&0.13&0.15\\
$\mu_{\Sigma^-}$&-0.18&-0.08&-0.09&-0.10&-0.11\\
$\mu_{\Sigma\Lambda}$&0.39&0.12&0.14&0.16&0.18\\
$\mu_{\Xi^-}$&-0.52&-0.10&-0.13&-0.16&-0.18\\
$\mu_{\Xi^0}$&-0.90&-0.17&-0.22&-0.26&-0.30
\end{tabular}
\caption{Given are the nonanalytic contributions to magnetic moments
in dimensional regularization and for various values of the cutoff
parameter $\Lambda$ in MeV.}
\end{center}
\end{table}

\section{Summary}

We have seen above that a significant component of the poor convergence
found in previous calculations in
SU(3) baryon chiral perturbation theory is due to the inclusion of
large and spurious short-distance contributions when loop processes
are regularized dimensionally. The use of a momentum space cutoff
keeps only the long distance portion of the loops and leads to an
improved behavior.  Indeed although we have formulated our discussion
in terms of merely a different sort of regularization procedure
within the general framework of chiral perturbation theory, it is
interesting to note that our results are quite consistent with the sort
of SU(3) breaking effects found in chiral confinement models such as
the cloudy bag, when the effects of kaon and/or eta loops are
isolated\cite{ClB}.

We might ask why baryon chiral perturbation theory has this problem
while mesonic chiral theories do not. Most applications in mesons
work perfectly well using dimensional regularization. At first sight
one might argue that the separation scale in baryons corresponds to
lower energies because the physical size of baryons is larger than
mesons. While this is a true statement, it does not really answer
the question, since the baryon problem surfaces entirely within
the point particle theory. For some reason, given the same meson
masses, the loop corrections are larger in the baryonic point particle
theory compared to a mesonic point particle theory. This feature can
perhaps be blamed on the baryon propagator in the loop integral which,
being linear in the momentum, suppresses high momentum contributions
less than a corresponding quadratic mesonic propagator. However, the
existence of the problem is beyond doubt, given the troubles discussed
in the introduction. Fortunately, we do not as a consequence have to
abandon all such chiral calculations---a revised regularization scheme
seems capable of resolving the problem.

The simplicity that underlies baryon physics is more evident when
chiral loops are calculated with a long-distance regularization. In this
context, we hope that baryon chiral perturbation theory will
become more phenomenologically useful. One can hopefully now use the chiral
calculations in order to provide a model independent description of the very
long distance physics, and this can be a welcome addition to our
techniques for describing the low energy phenomenology of baryons.


\begin{thebibliography}{99}
\bibitem{GL} J. Gasser and H. Leutwyler, Ann. Phys. (NY) {\bf 158}, 142 (1984);
Nucl. Phys. {\bf B250}, 465 (1985).
\bibitem{HBCPT} J. Gasser, M.E. Sainio and A. Svarc, Nucl. Phys. {\bf B307}, 
 779 (1988); E. Jenkins and A.V. Manohar, Phys. Lett. {\bf B281}, 336 (1992);
for a comprehensive review, see: V. Bernard, N. Kaiser and U.G. Meissner, 
Int. J. Mod. Phys. {\bf E4}, 193 (1995).
\bibitem{GMO} M. Gell-Mann in M. Gell-Mann and Y. Ne'eman, {\bf The Eightfold
Way},
Benjamin, New York (1962) and Phys. Rev. {\bf 125}, 1067 (1962), S. Okubo,
Prog. Theo.
Phys. {\bf 27}, 949 (1962).
\bibitem{CHIM} V. Bernard, N. Kaiser and U.G. Meissner, Z. Phys. {\bf C60},
111  (1993).
\bibitem{BM} B. Borasoy and U.G. Meissner, Phys. Lett. {\bf B365}, 285 (1996);
Ann. Phys; P. Langacker and H. Pagels, Phys. Rev. {\bf D8}, 4595 (1975).
(NY) {\bf 254}, 192 (1997).
\bibitem{DHS} P. G. Ratcliffe, Phys. Lett. {\bf B242}, 271 (1990),
ibid. {\bf B365}, 383 (1996); M. Roos Phys. Lett. {\bf B246}, 179 (1990).
\bibitem{BSW} J. Bijnens, H. Sonoda, and M.B. Wise, Nucl. Phys. {\bf B261}, 185
(1985).
\bibitem{LS} B.W. Lee, Phys. Rev. Lett. {\bf 12},83 (1964) ;
H. Sugawara, Prog. Theor. Phys. {\bf 31},213 (1964). 
\bibitem{BH} B. Borasoy and B.R. Holstein, hep-ph/9805340, to be
published in the Eur. Phys. J. C.
\bibitem{CP} D.G. Caldi and H. Pagels, Phys.Rev.  {\bf D10}, 3739 (1974). 
\bibitem{MS} U.G. Meissner and S. Steininger, Nucl. Phys. {\bf B499}, 349
(1997).
\bibitem{DH} J.F. Donoghue and B.R. Holstein, hep-ph/9803312, to be
published in Physics Letters. 
\bibitem{B} B. Borasoy, ``Baryon masses and
sigma terms,'' hep-ph/9807453.
\bibitem{B1} B. Borasoy, UMass preprint (1998) ``Baryon axial currents'' .
\bibitem{Gerstein} I. Gerstein, R. Jackiw, B.W. Lee, and S. Weinberg,
Phys. Rev. {\bf D3}, 2486 (1971).
\bibitem{ClB} See, {\it e.g.} T. Yamaguchi et al., Nucl. Phys. {\bf A500}, 429
(1989);
K. Kubodera et al., Nucl. Phys. {\bf A439}, 695 (1985); S. Theberge et al.,
Phys. Rev.
{\bf D22}, 2838 (1980); A.W. Thomas, J. Phys {\bf G7}, L283 (1981).

\end{thebibliography}
\end{document}